\documentclass[prb,reprint,10pt,superscriptaddress,nobibnotes]{revtex4-1}
\usepackage[utf8]{inputenc}
\usepackage[sort&compress]{natbib}
\usepackage{siunitx}
\usepackage[margin=30mm]{geometry}
\usepackage{amsmath,amssymb}
\usepackage{graphicx}
\usepackage{booktabs}
\usepackage{tabularx}
\newcolumntype{Y}{>{\centering\arraybackslash}X}
\DeclareSIUnit\Rydberg{Ry}

\begin{document}

\title{Revealing electronic correlations in YNi$_2$B$_2$C using photoemission spectroscopy}

\author{Aki Pulkkinen}
\affiliation{New Technologies-Research Center, University of West Bohemia, 30614 Pilsen, Czech Republic}
\affiliation{Département de Physique and Fribourg Center for Nanomaterials, Université de Fribourg, CH-1700 Fribourg, Switzerland}

\author{Geoffroy Kremer}
\affiliation{Département de Physique and Fribourg Center for Nanomaterials, Université de Fribourg, CH-1700 Fribourg, Switzerland}
\affiliation{Institut Jean Lamour, UMR 7198, CNRS-Université de Lorraine, Campus ARTEM, 2 allée André Guinier, BP 50840, 54011 Nancy, France}

\author{Vladimir N. Strocov}
\affiliation{Paul Scherrer Institut, Swiss Light Source, 5232 Villigen PSI, Switzerland}

\author{Frank Weber}
\affiliation{Institute for Quantum Materials and Technologies, Karlsruhe Institute of Technology, 76021, Karlsruhe, Germany}

\author{Ján Minár}
\email{jminar@ntc.zcu.cz}
\affiliation{New Technologies Research Center, University of West Bohemia, 30614 Pilsen, Czech Republic}

\author{Claude Monney}
\email{claude.monney@unifr.ch}
\affiliation{Département de Physique and Fribourg Center for Nanomaterials, Université de Fribourg, CH-1700 Fribourg, Switzerland}

\begin{abstract}
We present a combined density functional theory (DFT), one-step model of photoemission, and soft x-ray angle-resolved photoemission spectroscopy (SX-ARPES) study of the electronic structure of the quaternary borocarbide superconductor YNi$_2$B$_2$C. Our analysis reveals the presence of moderate electronic correlations beyond the semilocal DFT within the generalized gradient approximation. We show that DFT and the full potential Korringa-Kohn-Rostoker method combined with the dynamical mean field theory (DFT+DMFT) with average Coulomb interaction $U=\SI{3.0}{\electronvolt}$ and the exchange energy $J=\SI{0.9}{\electronvolt}$ applied to the Ni $d$-states are necessary for reproducing the experimentally observed SX-ARPES spectra.

\end{abstract}

\maketitle

\section{Introduction}

YNi$_2$B$_2$C is an intermetallic borocarbide superconductor~\cite{cava1994} in the family RNi$_2$B$_2$C (R $=$ Y, Lu, Tm, Er, Ho), with superconducting transition temperature $T_{\rm C} = \SI{15.6}{\kelvin}$. Identified early as a conventional phonon-mediated $s$-wave superconductor~\cite{mattheiss1994,mattheiss1994_2,pickett1994,lee1994,lawrie1995}, YNi$_2$B$_2$C was later found to show signs of large anisotropy of the superconducting gap~\cite{nohara1997,nohara1999,nohara2000,boaknin2001,izawa2001,izawa2002,maki2002,martinez2003,watanabe2004,weber2008,baba2010,kawamura2017} and strong electron-phonon interactions near the superconducting transition~\cite{kurzhals2022} stimulating a theoretical mechanism proposed by Kontani~\cite{kontani2004}.
Very recently, an improved theoretical approach for the calculations of the critical temperature of conventional superconductors was proposed~\cite{pellegrini2022}, but leads only to a moderate improvement in the calculated critical temperature for the specific case of YNi$_2$B$_2$C.
The open questions about the superconductive phase and the nature of the superconducting gap anisotropy might relate to correlations in the electronic structure of YNi$_2$B$_2$C that have not been assessed so far.

Identifying electronic correlations in real materials is a delicate issue, that can be achieved by comparing the experimental spectroscopic data to \textit{ab-initio} results.  However, being many-body interactions by nature, electronic correlations pose a challenge for a proper theoretical treatment within density functional theory (DFT)~\cite{morisanchez2008}. Semilocal density functional approximations, such as the popular generalized gradient approximation (GGA), treat electrons as independent particles interacting via the mean field of the electron density. While vastly successful in predicting atomic structures, DFT-GGA is only able to describe a part of the correlation effects. This is particularly important in transition metals and lanthanides with partially filled $d$- and $f$-states, where the incomplete treatment of local (on-site) correlations in DFT-GGA often leads to discrepancies with experimentally observed energy bands. Local correlations are commonly applied by adding on-site Coulomb interaction $U$, either in a static (DFT+$U$), or frequency-dependent manner using dynamical mean field theory (DFT+DMFT)~\cite{georges1996}.

\begin{figure*}
\includegraphics[width=.99\textwidth]{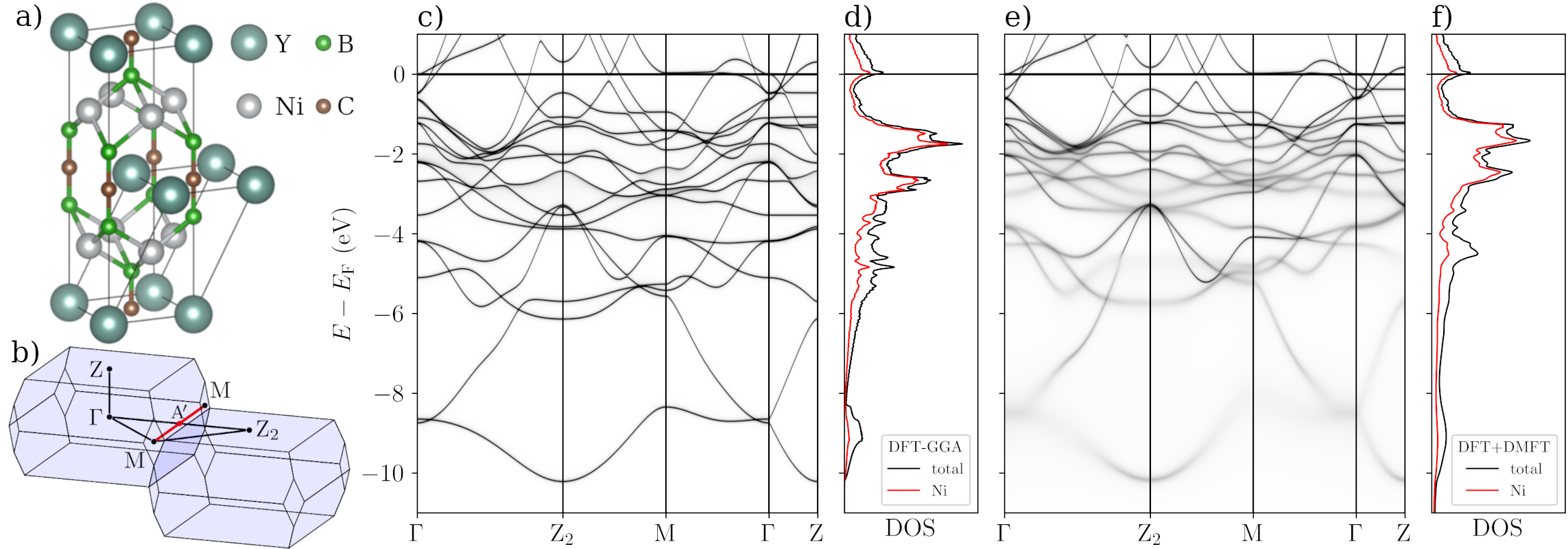}
\caption{\label{fig:crystalstructure}(a) Conventional and primitive unit cells of YNi$_2$B$_2$C. (b) Brillouin zone of YNi$_2$B$_2$C and the $k$-point path used in band structure calculations (black lines). The red line marks an additional path used in ARPES experiments and one-step model calculations. (c) Full potential KKR spectral function using the GGA. (d) DFT-GGA total DOS (black) and DOS projected on Ni $d$-states (red). (e) Full potential KKR spectral function within DFT+DMFT. (f) DFT+DMFT total DOS (black) and DOS projected on Ni $d$-states (red).}
\end{figure*}

Here we present an experimental and theoretical study of the electronic structure of YNi$_2$B$_2$C beyond semilocal DFT. By comparing soft x-ray angle-resolved photoemission spectroscopy (SX-ARPES) data with state-of-the-art one-step photoemission calculations using DFT+DMFT, we identify the presence of moderate electronic correlations in this compound and provide an estimate in terms of values for the Coulomb $U$ and exchange interaction $J$. We conclude that the influence of these electronic correlations on states close to $E_{\rm F}$ should be revisited by high-resolution ARPES and that their impact on superconductivity should also be considered in theoretical modelling.

\section{Methods}

\subsection{Experimental methods}
Soft x-ray ARPES experiments were performed at the SX-ARPES endstation~\cite{strocov2014} of the ADRESS beamline~\cite{strocov2010} of the Swiss Light Source using photons in the energy range $h\nu=$ \SIrange{680}{900}{\electronvolt}. Using photons in the soft x-ray energy range increases the photoelectron escape depth by a factor of 3-5 compared to the ultraviolet energy range. This leads to increased bulk sensitivity and improved resolution in the $k_z$ direction~\cite{strocov2012}. The experiments were performed for the (001) cleaved crystal surface at a temperature of \SI{20}{\kelvin} to reduce thermal motion that would negatively affect the momentum selectivity~\cite{braun2013}. Later, we refer to the momenta in the relative length units (\si{{r.l.u.}}), defined as $\SI{1}{{r.l.u.}} = 2\pi/a$, $2\pi/b$ and $2\pi/c$ in the $k_x$, $k_y$ and $k_z$ directions, respectively. 

To reveal details of the SX-ARPES spectra, the broad, inelastic background in the experimental spectra is subtracted. Details of the background subtraction procedure are given in the Supplementary Material.

\subsection{Theoretical methods}

The full potential Korringa-Kohn-Rostoker (FP-KKR) calculations were performed with the spin-polarized relativistic Korringa-Kohn-Rostoker (SPRKKR) package~\cite{ebert2021} within the generalized gradient approximation (GGA) using the Perdew-Burke-Ernzerhof (PBE) exchange-correlation functional and basis set truncated at $l_{\rm max}=4$. In the SPRKKR package, the relativistic phenomena are included at the level of the Dirac equation. 

The relativistic DFT+DMFT calculations were performed self-consistently with respect to the charge density and self-energy within the SPRKKR code, using the second order perturbative fluctuation exchange approximation (FLEX) solver~\cite{minar2005,minar2011}. The DFT+DMFT implementation follows the rotationally invariant LSDA+$U$ formulation of Liechtenstein~\emph{et al.}~\cite{liechtenstein1995}. The Slater integral $F^0$ is set equal to $U$, and the $F^2$ and $F^4$ are connected to $J$ by the relations $J = (F^2 + F^4)/14$ and $F^4/F^2=0.625$. The DMFT parameters applied to Ni $d$-states ($U=\SI{3.0}{\electronvolt}$, $J=\SI{0.9}{\electronvolt}$ and $T=\SI{400}{\kelvin}$) are the same as determined for bulk Ni~\cite{minar2005}.

The one-step model of photoemission~\cite{pendry1976} used for ARPES simulations is implemented in the SPRKKR code~\cite{braun2018}. The one-step model is based on Green's function and multiple scattering formalism and accounts for matrix element effects such as photon energy and polarization, experimental geometry, final state and surface effects. Therefore, the one-step model calculations allow direct comparisons to experimental ARPES spectra. 

The matrix elements were calculated with the full potential formalism inside the muffin-tin spheres. The experimental geometry of the ADRESS beamline SX-ARPES endstation~\cite{strocov2014}, with incoming photon direction \SI{70}{\degree} with respect to surface normal, was used in the ARPES simulations. Lifetime effects of the initial and final states were simulated by a constant imaginary part of the potential, \SI{0.05}{\electronvolt} and \SI{2.0}{\electronvolt}, respectively. To ensure that the momentum perpendicular to the surface, $k_z$, sampled in the one-step model calculations matches the experiment, we calculate a cut in the $(k_x,k_z)$ plane by varying the photon energy. The results are presented in the Supplementary Material. 

In all calculations, the structural parameters were fixed to experimental values. The crystal structure is body-centered tetragonal (space group $I4/mmm$, no. 139, see Fig.~\ref{fig:crystalstructure}~a)) with lattice parameters $a = \SI{3.526}{\angstrom}$ and $c = \SI{10.542}{\angstrom}$. The structure consists of layers of yttrium (at Wyckoff position $2b$) and carbon (at $2a$), and layers of nickel (at $4d$) surrounded by boron in distorted tetrahedral coordination (at $4e$, $z_{\rm B} = 0.1409$). The muffin-tin radii of the atom sites were set to \SI{1.746}{\angstrom} for Y, \SI{1.243}{\angstrom} for Ni, and \SI{0.741}{\angstrom} for B and C.

\section{Results}
\subsection{Ground state electronic structure}
In order to prepare the basis for studying the complex electronic band structure of YNi$_2$B$_2$C, we begin by looking at the bands calculated within DFT-GGA using the full potential KKR method (Fig.~\ref{fig:crystalstructure}~c)). The relatively large unit cell with 4 atomic species produces a complex band structure with 18 bands contributing to the occupied states down to \SI{-11}{\electronvolt}. The most prominent feature are the Ni $d$-bands that almost exclusively form the states between the Fermi level and \SI{-4}{\electronvolt}. The majority of the band character between \SI{-6}{\electronvolt} and \SI{-11}{\electronvolt} comes from B and C.

However, as we will see below with SX-ARPES data, only a few bands appear with a significant photoemission spectral weight within \SI{3}{\electronvolt} below the Fermi level. Therefore, to facilitate the comparison of the calculated band structure to the SX-ARPES spectra, we employ the one-step model of photoemission based on the multiple scattering KKR method. To this end, the prerequisite is a good quality set of scattering potentials calculated with the full potential KKR method, which we validate against the FP-LAPW method (see Supplementary Material). The excellent agreement between the band structures obtained with the FP-KKR, FP-LAPW and previously published band structures~\cite{youn2002,yamauchi2004,weber2014,kawamura2017} ensures a reliable starting point for our one-step model simulations.

\subsection{SX-ARPES and one-step model of photoemission}
\begin{figure*}
\includegraphics[width=\textwidth]{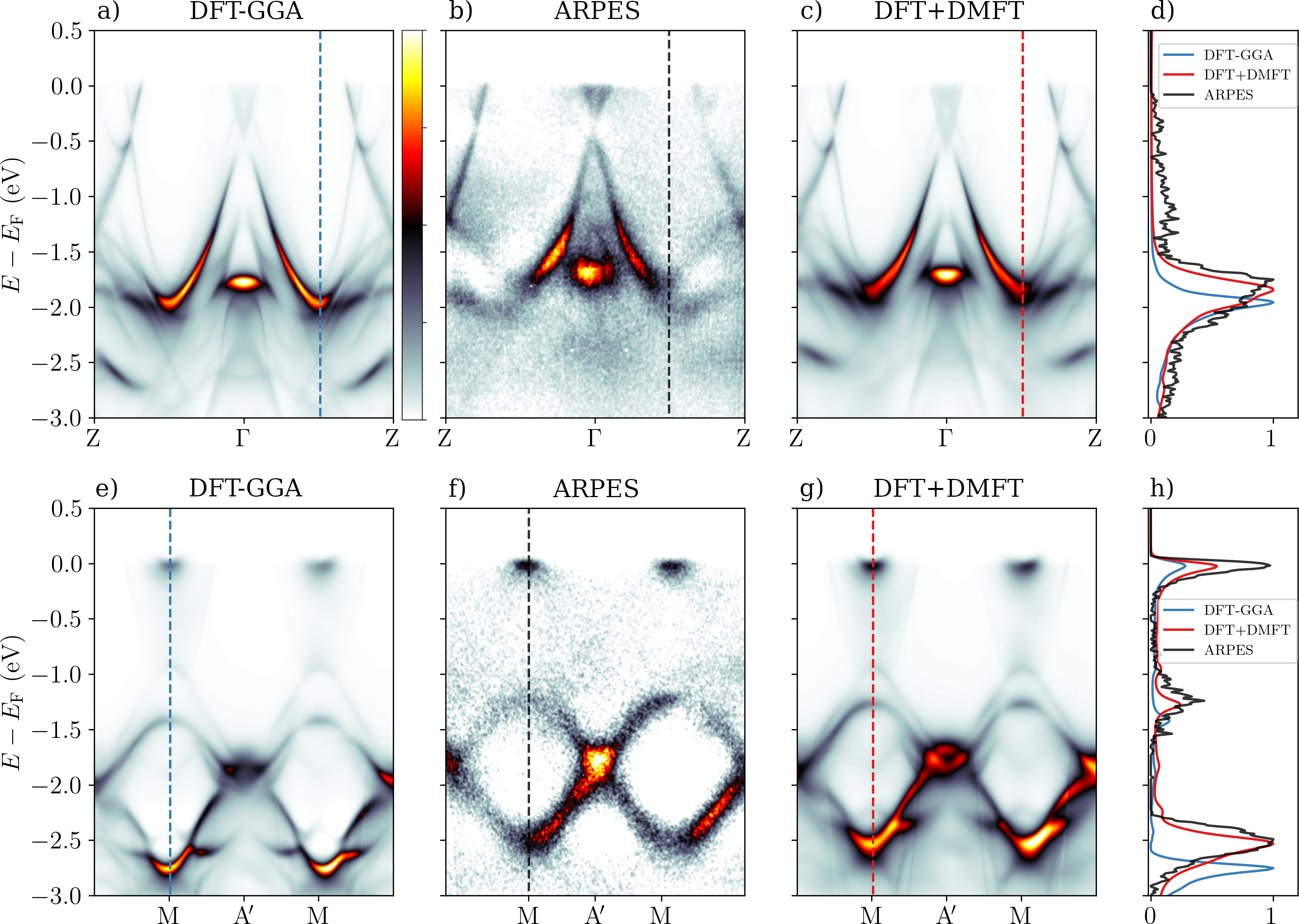}
\caption{\label{fig:kEmaps}Comparison of one-step model spectra along the directions  Z--$\Gamma$--Z (panels a)--c)) and M--$\rm{A}^{\!\prime}$--M (panels e)--g)) at $k_z = \SI{23}{{r.l.u.}}$ calculated with a) DFT-GGA and c) DFT+DMFT, and b) experiment using $p$-polarized light. Panels d) and h) show EDCs taken at the positions indicated by dashed lines in panels a)--c) and e)--g), respectively. The EDCs have been normalized so that the maximum value is 1.0.}
\end{figure*}

In Figs.~\ref{fig:kEmaps}~b) and f) we show SX-ARPES data acquired with $p$-polarization at a temperature of \SI{20}{\kelvin}. The photoemission intensity map along the path Z--$\Gamma$--Z at $k_z = \SI{23}{{r.l.u.}}$ ($h\nu=\SI{693}{\electronvolt}$) in Fig.~\ref{fig:kEmaps}~b) has a bright A-shaped feature centered at $\Gamma$, extending from \SI{-0.5}{\electronvolt} down to \SI{-2.0}{\electronvolt}. In addition, less intense bands are observed dispersing up from \SI{-1.2}{\electronvolt} at Z, along with electron pockets at $\Gamma$ and near Z. The one-step model spectrum calculated using the DFT-GGA in Fig.~\ref{fig:kEmaps}~a) captures well the main features and spectral weight distribution in the experimental spectrum in Fig.~\ref{fig:kEmaps}~b). However, closer inspection reveals that the DFT-GGA bands near \SI{-2.0}{\electronvolt} are located higher in binding energy than in the experiment. 

To access bands further away from $E_{\rm F}$, we investigate a photoemission intensity map along the direction M--${\rm A}^{\!\prime}$--M (see the BZ in Fig.~\ref{fig:crystalstructure}~b)) at $k_z = \SI{23}{{r.l.u.}}$ (Fig.~\ref{fig:kEmaps}~f)). The ARPES data shows small electron pockets at the M points near $E_{\rm F}$ and ring-like features composed of upward and downward dispersing bands between \SI{-1.4}{\electronvolt} and \SI{-2.5}{\electronvolt}. The comparison to DFT-GGA one-step model result in Fig.~\ref{fig:kEmaps}~e) shows even larger shifts in band positions at higher binding energy, even though the overall shape of the intensity map is similar to experiment. The discrepancies are not explained by a rigid shift of the bands, because the electron pockets are correctly positioned close to $E_{\rm F}$. Rather, it is a question of the well known delocalization problem caused by insufficient treatment of correlation effects in DFT-GGA~\cite{morisanchez2008}, which leads to band width overestimation of partially filled $d$- and $f$-bands. We will therefore assess the effect of dynamical correlations to the ARPES spectra by comparing the one-step model spectra obtained with DFT+DMFT to the DFT-GGA and experimental spectra. A direct comparison between one-step model spectra and full potential KKR Bloch spectral function within DFT can be found in the Supplementary Material.

For this purpose we draw attention to the DOS in Fig.~\ref{fig:crystalstructure}~d). The majority of the states in the energy range \SI{-4}{\electronvolt} to $E_{\rm F}$ have Ni $d$-character. In addition, the Ni sublattice in YNi$_2$B$_2$C crystal structure (Fig.~\ref{fig:crystalstructure}~a)) is strongly reminiscent of the structure of bulk Ni. Since the YNi$_2$B$_2$C lattice parameter $a$ is very close to the lattice parameter of face centered cubic (fcc) Ni ($a_{\rm Ni} = \SI{3.524}{\angstrom}$), the geometry of the square Ni planes YNi$_2$B$_2$C is essentially identical to the $\left\{001\right\}$ planes of fcc-Ni. Therefore, to reduce the delocalization problem in the electronic structure, we choose to apply the DFT+DMFT to the $d$-states of Ni in YNi$_2$B$_2$C with the same parameters ($U=\SI{3.0}{\electronvolt}$, $J=\SI{0.9}{\electronvolt}$) as already determined for bulk fcc-Ni~\cite{minar2005}.

The DFT+DMFT band structure and DOS are presented in Figs.~\ref{fig:crystalstructure}~e) and f). As DMFT is applied to the Ni $d$-states, the Ni bands between \SI{-1}{\electronvolt} and \SI{-4}{\electronvolt} shift towards the Fermi energy, reducing the $d$-band width. Most of the bands with mixed Ni and B character at energies from \SI{-4}{\electronvolt} to \SI{-10}{\electronvolt} do not shift significantly, but become very diffuse due to the imaginary part of the self-energy $\mathop{\mathrm{Im}} \Sigma$. An exception is the band near the Z point with strongly mixed Ni and B character that shifts towards $E_{\rm F}$ by about \SI{1.0}{\electronvolt}.

The one-step model spectra calculated with DFT+DMFT are presented in Fig.~\ref{fig:kEmaps}~c) and g). The electron pockets close to $E_{\rm F}$ at the M points in the M--${\rm A}^{\!\prime}$--M spectra are not shifted, but the bands at higher binding energy are shifted towards $E_{\rm F}$, leading to a very good agreement with the experimental spectra. To quantify the improvement brought by DFT+DMFT, we show energy distribution curves (EDCs) in Fig.~\ref{fig:kEmaps}~d) and h) cut at in-plane momenta shown by dashed lines in Fig.~\ref{fig:kEmaps}~a) - c) and e) - g). Along the Z--$\Gamma$--Z direction (Fig.~\ref{fig:kEmaps}~d)), the DFT-GGA peak is \SI{170}{\milli\electronvolt} lower than the corresponding experimental peak, and DFT+DMFT shifts the peak up by \SI{110}{\milli\electronvolt} to a very good agreement with ARPES. In addition, the broadening of the bands by the imaginary part $\mathop{\mathrm{Im}} \Sigma$ of the self-energy leads to a wider peak shape in DFT+DMFT that corresponds more closely to the experiment. Along the M--${\rm A}^{\!\prime}$--M direction (Fig.~\ref{fig:kEmaps}~h)), the bottom of the band at M is shifted up by \SI{220}{\milli\electronvolt} and the top of the band by \SI{150}{\milli\electronvolt}, bringing the peaks very close to the band positions in the experiment. 

\begin{figure*}
\includegraphics[width=\textwidth]{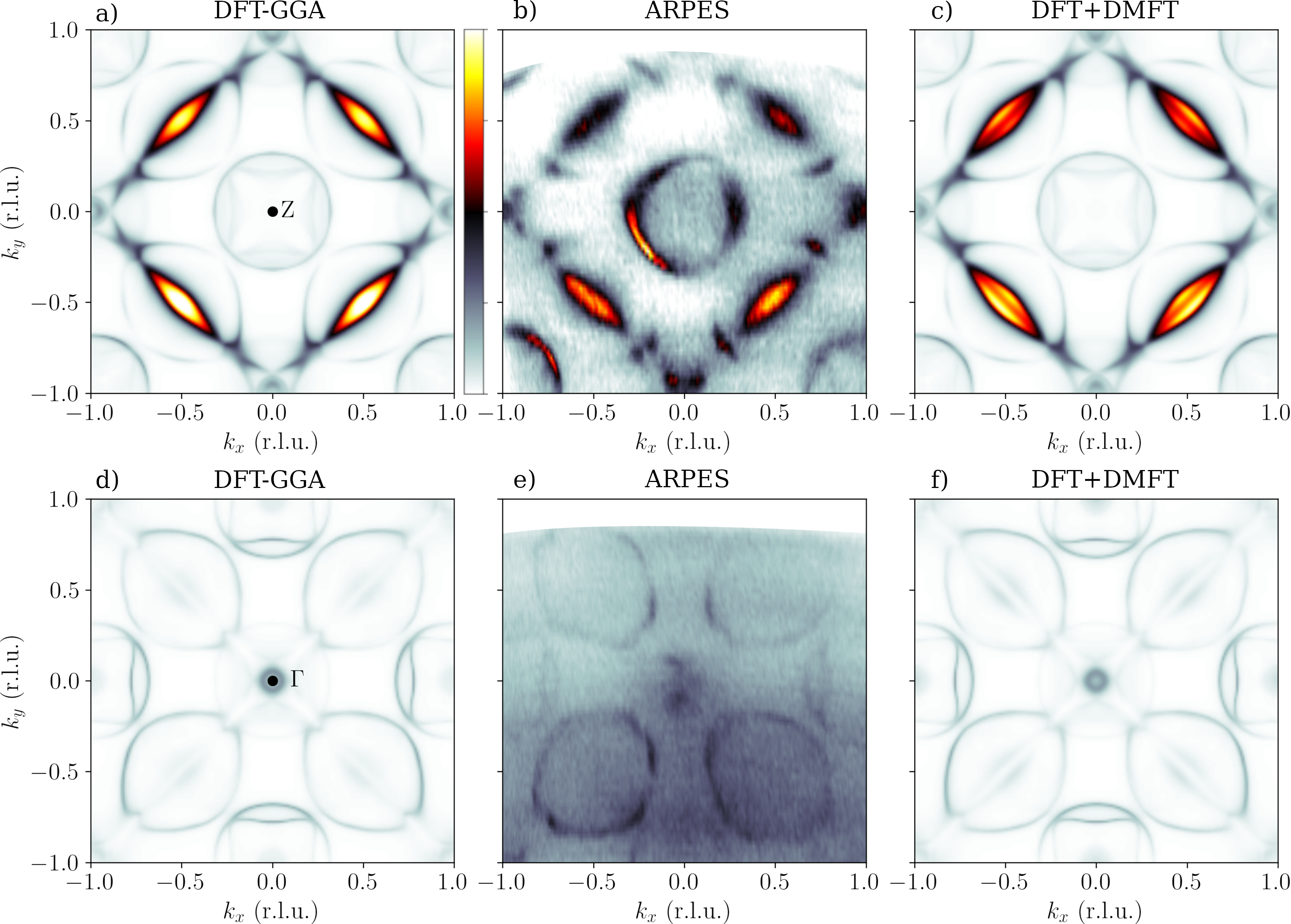}
\caption{\label{fig:FS_23_24}Comparison of calculated and SX-ARPES Fermi surface maps at $k_z=\SI{23}{{r.l.u.}}$ (a) - c)) and $k_z=\SI{24}{{r.l.u.}}$ (d) - f)). The high-symmetry point at the center of the map is indicated in panels a) and d).}
\end{figure*}

Finally, it is interesting to investigate if DFT+DMFT has an effect on the Fermi surface (FS) of YNi$_2$B$_2$C, as DFT band structure calculations have shown that the FS is well described already at the level of local density approximation (LDA)~\cite{weber2014,kurzhals2022}. One-step model FS maps are compared to the SX-ARPES FS maps $k_z=\SI{23}{{r.l.u.}}$ ($h\nu=\SI{693}{\electronvolt}$) and $k_z=\SI{24}{{r.l.u.}}$ ($h\nu=\SI{760}{\electronvolt}$) in Fig.~\ref{fig:FS_23_24}. Note that the contribution of the photon momentum $k_{\rm phot} = h\nu/c$ has been subtracted so that the electron momentum $(k_x,k_y)=(0,0)$ is at the center of the map. The experimental spectral weight distribution is very well reproduced by both DFT-GGA and DFT+DMFT, except for at the center of the map, where the experimental spectra has stronger spectral weight than either of the one-step model maps. The DFT-GGA and DFT+DMFT maps are remarkably similar, apart from a small increase of spectral weight at the Z points in the DFT+DMFT map in Fig.~\ref{fig:FS_23_24}~f)), which we attribute to a downward shift of an Y-character band above $E_{\rm F}$.

Our present work therefore evidences the presence of moderate, but substantial electronic correlations with $U=\SI{3.0}{\electronvolt}$ and $J=\SI{0.9}{\electronvolt}$ in the Ni $d$-states of YNi$_2$B$_2$C from an analysis of its electronic structure on a few-eV energy scale, indicating energy shifts of the order of \SI{200}{\milli\electronvolt} within \SI{3}{\electronvolt} below the $E_{\rm F}$. Now the question arises whether these electronic correlations affect substantially states close to $E_{\rm F}$ that are relevant for superconductivity and for anomalously broad phonon lineshape at low temperatures\cite{kurzhals2022}.
In the former case, Baba and coworkers identified with very high resolution ARPES the presence of superconductivity-induced band gaps in different Fermi surface pockets at a wave vector $k_z$ value close to the Z point (which corresponds to $k_z = \SI{23}{{r.l.u.}}$ in our present notation)\cite{baba2010}. The band gaps were shown to be particularly large for the pockets close to $(k_x,k_y)=(0, 0)$. In the latter case, the electron-momentum dependence of electron-phonon coupling is substantially boosted by the presence of square-like pockets centered around $(k_x,k_y) = (0.5, 0.5)$ in the same $k_z$ plane.

\begin{figure*}[!ht]
\includegraphics[width=0.5\textwidth]{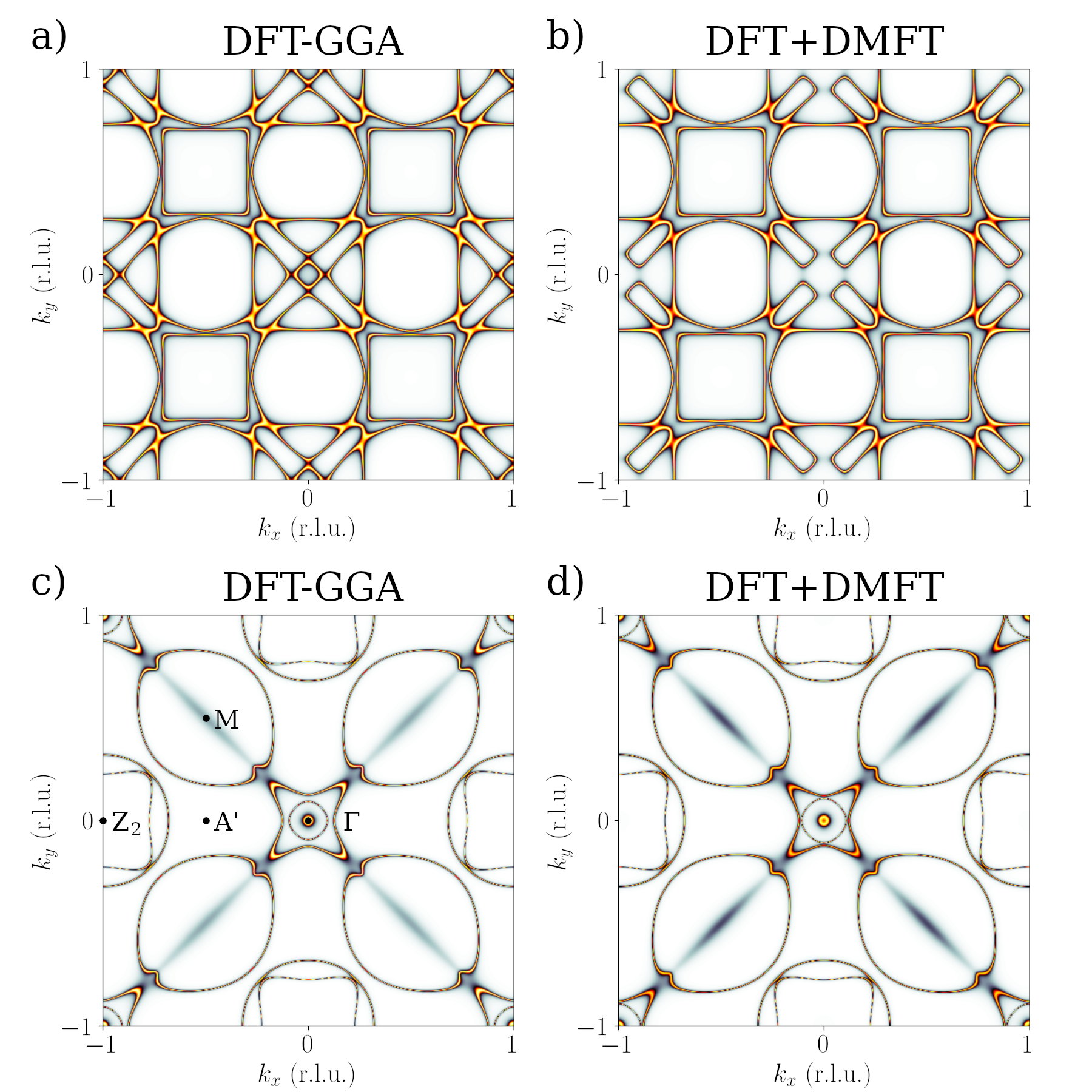}
\caption{\label{fig:FS_BSF}Full potential KKR spectral function Fermi surface cuts at $k_z=\SI{0.5}{{r.l.u.}}$ (a) - b)) and $k_z=\SI{0.0}{{r.l.u.}}$ (c) - d)).}
\end{figure*}

We have therefore calculated Fermi surfaces with DFT-GGA and DFT+DMFT ($U=\SI{3.0}{\electronvolt}$, $J=\SI{0.9}{\electronvolt}$), to assess the influence of electronic correlations beyond DFT-GGA (see Fig.~\ref{fig:FS_BSF}) on these different contributions. A broadening of \SI{1}{\milli\Rydberg} was used, meaning an integration over \SI{14}{\milli\electronvolt} around $E_{\rm F}$.

Our calculations (Fig.~\ref{fig:FS_BSF}~a) and Fig.~\ref{fig:FS_BSF}~b)) show that, while moderate correlations do not affect the square-like pockets for $k_z=\SI{0.5}{{r.l.u.}}$, they significantly modify the Fermi surface contributions near $(k_x,k_y)=(0,0)$. As a comparison, we also show the FS calculated for $k_z=\SI{0.0}{{r.l.u.}}$ (Fig.~\ref{fig:FS_BSF}~c) and Fig.~\ref{fig:FS_BSF}~d)), for which similar trends are observed, but of smaller amplitude. These comparisons therefore support the idea that electronic correlations do not influence the electron-momentum dependence of electron-phonon coupling, as described by Kurzhals \emph{et al.}~\cite{kurzhals2022} However, they suggest that the electronic states in the vicinity of the Fermi level that become gapped upon superconductivity are substantially modified by electronic correlations. Note that the ARPES measurements and one-step model calculations at integer $k_z$ (\SI{23}{{r.l.u.}} and \SI{24}{{r.l.u.}}) in Fig.~\ref{fig:FS_23_24} do not sample the square pockets which exist at half-integer $k_z$.

Our SX-ARPES data do not allow to resolve fine details in electronic dispersions on an energy scale of \SIrange{10}{30}{\milli\electronvolt} that is relevant for superconductivity and renormalization of phonon lineshapes and we are unable to make a one-to-one comparison with experimental data. Additional low-energy ARPES data with sufficient high resolution, as in the study of Baba and coworkers, are therefore necessary to evaluate if our theoretical prediction is correct. In particular, it would be very helpful to map the electronic dispersions close to $E_{\rm F}$ for temperatures above and below the critical temperature of superconductivity.

What would be the consequence of moderate electronic correlations on Ni $d$-states? YNi$_2$B$_2$C has initially attracted attention as a potential Ni-based conventional superconductor with high critical temperature due to a phonon-mediated pairing interaction inducing a superconducting gap of about \num{2}-\SI{3}{\milli\electronvolt}\cite{mattheiss1994,mattheiss1994_2,pickett1994,lee1994,lawrie1995,cheon1999}. However, a large anisotropy of the superconducting gap was later discovered, challenging the possibility of conventional superconductivity~\cite{nohara1997,boaknin2001}. Recently, a theoretical approach based on the Eliashberg theory of superconductivity and including ab-initio static Coulomb interaction was used to calculate the critical temperature $T_c$ of several conventional superconductors. It arrived this way to a value of $T_c$  for the case of YNi2B2C about \SI{25}{\percent} lower than the experimental value. Interestingly, a very recent study reported an ab-initio calculation of the dynamically screened electron-electron Coulomb interaction leading to moderate electronic correlations with $U=\SI{3}{\electronvolt}$. The same authors calculated a value of $T_c$ even closer to the experimental one~\cite{christiansson2024}. In that context, our work brings an experimental confirmation that moderate electronic correlations are present in YNi$_2$B$_2$C. This is particularly relevant in view of the work of Kontani~\cite{kontani2004}. In the proposed theory accounting for the anisotropic $s$-wave superconductivity, antiferromagnetic fluctuations originating from an on-site Coulomb interaction $U$ were also treated within a FLEX-type
approximation and play a central role. It would therefore be interesting to assess the necessary value for the magnetic interaction term in view of our estimation for $U$ and $J$.

\section{Conclusions}
In this work, we compare one-step model calculations of photoemission using DMFT to SX-ARPES data and reveal the presence of moderate electronic correlations on the Ni $d$-states.
In a recent work~\cite{kurzhals2022}, it was demonstrated that electron-phonon coupling in YNi$_2$B$_2$C is strongly enhanced for specific values of electron momentum. Extrapolating from our SX-ARPES data and calculated Fermi surfaces, we anticipate that these moderate electronic correlations affect significantly the electronic states that participate in superconductivity and we propose new very high resolution ARPES studies to assess this conjecture. 
These new results, namely the moderate electronic correlations and momentum-dependent electron-phonon coupling, provide new fundamental input for models of superconductivity in YNi$_2$B$_2$C.

\begin{acknowledgements}
We thank our collaborators Philipp Kurzhals, Thomas Jaouen, Christopher W. Nicholson, Peter Nagel, Maxime Rumo and Björn Salzmann for their support during the ARPES measurements. A.P. and C.M. acknowledge support from the Swiss National Science Foundation Grant No. P00P2{\_}170597. J.M. and A.P.
would like to thank the QM4ST project with Reg. No.
CZ.02.01.01{/}00{/}22{\_}008{/}0004572, cofunded by the ERDF as
part of the MŠMT.
\end{acknowledgements}

\bibliography{bibliography.bib}

\end{document}